\documentclass[12pt]{article}
\usepackage{graphicx}
\usepackage{amsfonts}
\usepackage{amsmath}
\usepackage{cite}

\def\beq{\begin{equation}}
\def\eeq{\end{equation}}
\def\nn{\nonumber}
\def\bea{\begin{eqnarray}}
\def\eea{\end{eqnarray}}
\def\ba{\begin{array}}                  
\def\ea{\end{array}}

\newcommand{\br}{{\rm BR}}

\usepackage[latin1]{inputenc}
\usepackage[T1]{fontenc}
\newcommand{\tb}{\tan \beta}
\newcommand{\FUTA}{{\bf FUTA}}
\newcommand{\FUTB}{{\bf FUTB}}

\begin{document}


\begin{center}

{\Large\bf \boldmath Finiteness and the Higgs mass prediction} 

\vspace*{6mm}

{S.~Heinemeyer$^{a}$, M~Mondrag\'on$^{b}$ and G.~Zoupanos$^{c}$ }\\      

{\small \it $^a$ Instituto de Física de Cantabria (CSIC-UC),\\
Edificio Juan Jorda,
Avda. de Los Castros s/n\\
39005 Santander, Spain \\      
$^b$ Inst.~de Física, Universidad~Nacional Aut\'onoma de M\'exico,\\
Apdo. Postal 20-364, M\'exico 01000 D.F., M\'exico
\\
$^c$ Physics Department, National Technical University of Athens,\\
Zografou Campus: Heroon Polytechniou 9,\\
15780 Zografou, Athens, Greece }

\end{center}

\vspace*{6mm}


\begin{abstract}

  Finite Unified Theories (FUTs) are $N=1$ supersymmetric Grand
  Unified Theories (GUTs) which can be made finite to all-loop orders,
  leading to a drastic reduction in the number of free parameters.  By
  confronting the predictions of $SU(5)$ FUTs with the top and bottom
  quark masses we are able to discriminate among different models.
  Including further low-energy phenomenology constraints, such as $B$
  physics observables, the bound on the SM Higgs mass and the cold dark
  matter density, we derive predictions for the lightest Higgs boson
  mass and the sparticle spectrum.

\end{abstract}

\vspace*{6mm}

\section{Introduction}

Finite Unified Theories (FUTs) are $N=1$ supersymmetric Grand Unified
Theories (GUTs) which can be made finite to all-loop orders, including
the soft supersymmetry breaking sector.  FUTs have always attracted
interest for their intriguing mathematical properties and their
predictive power.  To construct GUTs with reduced independent
parameters \cite{Kubo:1994bj,zoup-zim1} one has to search for
renormalization group invariant (RGI) relations holding below the
Planck scale, which in turn are preserved down to the GUT scale.  This
programme, called Gauge--Yukawa unification scheme, applied in the
dimensionless couplings of supersymmetric GUTs, such as gauge and
Yukawa couplings, had already noticeable successes by predicting
correctly, among others, the top quark mass in the finite $SU(5)$ GUTs
\cite{zoup-finite1,zoup-kkmz1}.  An impressive aspect of the RGI
relations is that one can guarantee their validity to all-orders in
perturbation theory by studying the uniqueness of the resulting
relations at one-loop, as was proven in the early days of the
programme of {\it reduction of couplings} \cite{zoup-zim1}. Even more
remarkable is the fact that it is possible to find RGI relations among
couplings that guarantee finiteness to all-orders in perturbation
theory \cite{zoup-lucchesi1,zoup-ermushev1}. 

The search for RGI relations and finiteness has been extended to the
soft supersymmetry breaking sector (SSB) of these theories
\cite{zoup-kkk1,zoup-kkz,Jack:1994kd,zoup-kmz2,jack4} which involves
parameters of dimension one and two.  An interesting observation at
the time was that in $N = 1$ Gauge--Yukawa unified
theories there exists a RGI sum rule for the soft scalar masses at
lower orders; at one-loop for the non-finite case \cite{zoup-kkk1} and
at two-loops for the finite case \cite{zoup-kkmz1}. The sum rule,
although introducing more parameters in the theory, manages to overcome
the problems that the universality condition for the soft scalar
masses in FUTs had: that the lightest supersymmetric particle is
charged, namely, the stau $\tilde\tau$ (although this is not
necessarily a problem, as we will see), that it is difficult to comply
with the attractive radiative electroweak symmetry breaking, and,
worst of all, that the universal soft scalar masses can lead to charge
and/or colour breaking minima deeper than the standard
vacuum. Moreover, it was proven \cite{zoup-kkz} that the sum rule for
the soft scalar massses is RGI to all-orders for both the general as
well as for the finite case.  Finally the exact $\beta$-function for
the soft scalar masses in the Novikov-Shifman-Vainstein-Zakharov
(NSVZ) scheme \cite{zoup-novikov1} for the softly broken
supersymmetric QCD has been obtained \cite{zoup-kkz}.  Eventually, the
full theories can be made all-loop finite and, with use of the sum
rule, their predictive power is extended to the Higgs sector and the
SUSY spectrum.  Thus, we are now in a position to study the spectrum
of the full finite $SU(5)$ models in terms of few free parameters with
emphasis on the predictions for the masses of the lightest Higgs and
LSP and on the constraints imposed by low-energy phenomenology
observables.

\section{ FINITE UNIFIED THEORIES}

Finiteness can be understood by considering a chiral, anomaly free,
$N=1$ globally supersymmetric
gauge theory based on a group G with gauge coupling
constant $g$. The
superpotential of the theory is given by
\begin{equation}
 W= \frac{1}{2}\,m^{ij} \,\Phi_{i}\,\Phi_{j}+
\frac{1}{6}\,C^{ijk} \,\Phi_{i}\,\Phi_{j}\,\Phi_{k}~, 
\label{1}
\end{equation}
where $m^{ij}$ (the mass terms) and $C^{ijk}$ (the Yukawa couplings) are
gauge invariant tensors and 
the matter field $\Phi_{i}$ transforms
according to the irreducible representation  $R_{i}$
of the gauge group $G$. 
All the one-loop $\beta$-functions of the theory
vanish if  the $\beta$-function of the gauge coupling $\beta_g^{(1)}$, and
the anomalous dimensions of the Yukawa couplings $\gamma_i^{j(1)}$, vanish, i.e.
\begin{equation}
\sum _i \ell (R_i) = 3 C_2(G) \,,~
\frac{1}{2}C_{ipq} C^{jpq} = 2\delta _i^j g^2  C_2(R_i)\ ,
\label{zoup-fini}
\end{equation}
where $\ell (R_i)$ is the Dynkin index of $R_i$, and $C_2(G)$ is the
quadratic Casimir invariant of the adjoint 
representation of $G$.

The conditions for finiteness for $N=1$ field theories with $SU(N)$ gauge
symmetry are discussed in \cite{Rajpoot:1984zq}, and the
analysis of the anomaly-freedom and no-charge renormalization
requirements for these theories can be found in \cite{Rajpoot:1985aq}. 
A very interesting result is that the conditions (\ref{zoup-fini}) are
necessary and sufficient for finiteness at
the two-loop level \cite{Parkes:1984dh}.
 
A powerful theorem \cite{zoup-lucchesi1} guarantees
the vanishing of the $\beta$-functions to all-orders in perturbation
theory.  This requires that, in addition to the
one-loop finiteness conditions (\ref{zoup-fini}),  
the Yukawa couplings are reduced
in favour of the gauge coupling.
Alternatively, similar results can be
obtained  \cite{zoup-ermushev1,zoup-strassler} using an analysis of
the all-loop NSVZ gauge beta-function \cite{zoup-novikov1}. 

In the soft breaking sector, it was found that RGI SSB scalar masses
in Gauge-Yukawa unified models satisfy a universal sum rule at
one-loop \cite{zoup-kkk1}. This result was generalized to two-loops
for finite theories \cite{zoup-kkmz1}, and then to all-loops for
general Gauge-Yukawa and finite unified theories \cite{zoup-kkz}.
Then the following soft scalar-mass sum rule is found
\cite{zoup-kkmz1}
\begin{equation}
\frac{(~m_{i}^{2}+m_{j}^{2}+m_{k}^{2}~)}{M M^{\dag}} =
1+\frac{g^2}{16 \pi^2}\,\Delta^{(2)}
+O(g^4)~
\label{zoup-sumr}
\end{equation}
for i, j, k with $\rho^{ijk}_{(0)} \neq 0$, where $\Delta^{(2)}$ is
the two-loop correction
\begin{equation}
\Delta^{(2)} =  -2\sum_{l} [(m^{2}_{l}/M M^{\dag})-(1/3)]~ \ell (R_l),
\label{5}
\end{equation}
$\Delta^{(2)}$ vanishes for the
universal choice, i.e.\ when all the soft scalar masses are the same at
the unification point.

A realistic two-loop finite $SU(5)$ model was presented in
\cite{Jones:1984qd}, and shortly afterwards the conditions for
finiteness in the soft susy breaking sector at one-loop
\cite{Jones:1984cu} were given.  Since these finite models have
usually an extended Higgs sector, in order to make them viable a
rotation of the Higgs sector was proposed \cite{Leon:1985jm}.  The
first all-loop finite theory was studied in \cite{zoup-finite1},
without taking into account the soft breaking terms.  Naturally, the
concept of finiteness was extended to the soft breaking sector, where
also one-loop finiteness implies two-loop finiteness
\cite{Jack:1994kd}, and then finiteness to all-loops in the soft
sector of realistic models was studied
\cite{Kazakov:1995cy,Kazakov:1997nf}, although the universality of the
soft breaking terms lead to a charged LSP. This fact was also noticed
in \cite{Yoshioka:1997yt}, where the inclusion of an extra parameter
in the Higgs sector was introduced to alleviate it. With the
derivation of the sum-rule in the soft supersymmetry breaking sector
and the proof that it can be made all-loop finite the construction of
all-loop phenomenologically viable finite models was made possible
\cite{zoup-kkz,zoup-kkmz1}.

 Here we will examine such all-loop Finite Unified theories
with $SU(5)$ gauge group, where the reduction of couplings has been
applied to the third generation of quarks and leptons.  An extension
to three families, and the generation of quark mixing angles and
masses in Finite Unified Theories has been addressed in
\cite{Babu:2002in}, where several examples are given. These
extensions are not considered here.  Realistic Finite Unified Theories
based on product gauge groups, where the finiteness implies three
generations of matter, have also been studied \cite{Ma:2004mi}.

The particle content of the models we will study consists of the
following supermultiplets: three ($\overline{\bf 5} + \bf{10}$),
needed for each of the three generations of quarks and leptons, four
($\overline{\bf 5} + {\bf 5}$) and one ${\bf 24}$ considered as Higgs
supermultiplets. 
When the gauge group of the finite GUT is broken the theory is no
longer finite, and we will assume that we are left with the MSSM.

Thus, a predictive Gauge-Yukawa unified $SU(5)$ model which is finite to all
orders, in addition to the requirements mentioned already, should also
have the following properties:
\begin{enumerate}
\item 
One-loop anomalous dimensions are diagonal,
i.e.,  $\gamma_{i}^{(1)\,j} \propto \delta^{j}_{i} $.
\item Three fermion generations, in the irreducible representations
  $\overline{\bf 5}_{i},{\bf 10}_i~(i=1,2,3)$, which obviously should
  not couple to the adjoint ${\bf 24}$.
\item The two Higgs doublets of the MSSM should mostly be made out of a
pair of Higgs quintet and anti-quintet, which couple to the third
generation.
\end{enumerate}

In the following we discuss two versions of the all-order finite
model.  The model of ref.~\cite{zoup-finite1}, which will be labeled
${\bf A}$, and a slight variation of this model (labeled ${\bf B}$),
which can also be obtained from the class of the models suggested in
ref.~\cite{Kazakov:1995cy} with a modification to suppress
non-diagonal anomalous dimensions.

The  superpotential which describes the two models 
takes the form \cite{zoup-finite1,zoup-kkmz1}
\bea
W &=& \sum_{i=1}^{3}\,[~\frac{1}{2}g_{i}^{u}
\,{\bf 10}_i{\bf 10}_i H_{i}+
g_{i}^{d}\,{\bf 10}_i \overline{\bf 5}_{i}\,
\overline{H}_{i}~] \nn \\\nn
&+&g_{23}^{u}\,{\bf 10}_2{\bf 10}_3 H_{4} 
  +g_{23}^{d}\,{\bf 10}_2 \overline{\bf 5}_{3}\,
\overline{H}_{4}+
g_{32}^{d}\,{\bf 10}_3 \overline{\bf 5}_{2}\,
\overline{H}_{4} \\
&+&\sum_{a=1}^{4}g_{a}^{f}\,H_{a}\, 
{\bf 24}\,\overline{H}_{a}+
\frac{g^{\lambda}}{3}\,({\bf 24})^3~,
\label{zoup-super}
\eea
where 
$H_{a}$ and $\overline{H}_{a}~~(a=1,\dots,4)$
stand for the Higgs quintets and anti-quintets.

The non-degenerate and isolated solutions to $\gamma^{(1)}_{i}=0$ for
the models $\{ {\bf A}~,~{\bf B} \}$ are: 
\bea (g_{1}^{u})^2
&=&\{\frac{8}{5},\frac{8}{5} \}g^2~, ~(g_{1}^{d})^2
=\{\frac{6}{5},\frac{6}{5}\}g^2~,~\nn\\
(g_{2}^{u})^2&=&(g_{3}^{u})^2=\{\frac{8}{5},\frac{4}{5}\}g^2~,\label{zoup-SOL5}\\
(g_{2}^{d})^2 &=&(g_{3}^{d})^2=\{\frac{6}{5},\frac{3}{5}\}g^2~,~\nn\\
(g_{23}^{u})^2 &=&\{0,\frac{4}{5}\}g^2~,~
(g_{23}^{d})^2=(g_{32}^{d})^2=\{0,\frac{3}{5}\}g^2~,
\nonumber\\
(g^{\lambda})^2 &=&\frac{15}{7}g^2~,~ (g_{2}^{f})^2
=(g_{3}^{f})^2=\{0,\frac{1}{2}\}g^2~,~ \nn\\
(g_{1}^{f})^2&=&0~,~
(g_{4}^{f})^2=\{1,0\}g^2~.\nonumber 
\eea 
According to the theorem of
ref.~\cite{zoup-lucchesi1} these models are finite to all orders.  After the
reduction of couplings the symmetry of $W$ is enhanced
\cite{zoup-finite1,zoup-kkmz1}.

The main difference of the models ${\bf A}$ and ${\bf B}$ is that
two pairs of Higgs quintets and anti-quintets couple to the ${\bf
  24}$ for ${\bf B}$ so that it is not necessary to mix them with
$H_{4}$ and $\overline{H}_{4}$ in order to achieve the triplet-doublet
splitting after the symmetry breaking of $SU(5)$.

In the dimensionful sector, the sum rule gives us the following
boundary conditions at the GUT scale \cite{zoup-kkmz1}:
\bea
m^{2}_{H_u}+
2  m^{2}_{{\bf 10}} &=&
m^{2}_{H_d}+ m^{2}_{\overline{{\bf 5}}}+
m^{2}_{{\bf 10}}=M^2~~\mbox{for}~~{\bf A} ~;\\
m^{2}_{H_u}+
2  m^{2}_{{\bf 10}} &=&M^2~,~
m^{2}_{H_d}-2m^{2}_{{\bf 10}}=-\frac{M^2}{3}~,~\nonumber\\
m^{2}_{\overline{{\bf 5}}}+
3m^{2}_{{\bf 10}}&=&\frac{4M^2}{3}~~~\mbox{for}~~{\bf B},
\eea
where we use as  free parameters 
$m_{\overline{{\bf 5}}}\equiv m_{\overline{{\bf 5}}_3}$ and 
$m_{{\bf 10}}\equiv m_{{\bf 10}_3}$
for the model ${\bf A}$, and 
$m_{{\bf 10}}\equiv m_{{\bf 10}_3}$  for ${\bf B}$, in addition to $M$.

\section{PREDICTIONS OF LOW ENERGY PARAMETERS}
 
Since the gauge symmetry is spontaneously broken below $M_{\rm GUT}$,
the finiteness conditions do not restrict the renormalization properties
at low energies, and all it remains are boundary conditions on the
gauge and Yukawa couplings (\ref{zoup-SOL5}), the $h=-MC$ relation,
and the soft scalar-mass sum rule (\ref{zoup-sumr}) at $M_{\rm GUT}$,
as applied in the two models.  Thus we examine the evolution of
these parameters according to their RGEs up
to two-loops for dimensionless parameters and at one-loop for
dimensionful ones with the relevant boundary conditions.  Below
$M_{\rm GUT}$ their evolution is assumed to be governed by the MSSM.
We further assume a unique supersymmetry breaking scale $M_{s}$ (which
we define as the geometric mean of the stop masses) and
therefore below that scale the  effective theory is just the SM.

We now present the comparison of the predictions of the two models
(\FUTA, \FUTB) with the experimental data, starting with the heavy
quark masses see ref.~\cite{Heinemeyer:2007tz} for more details.  For the
top quark pole mass we used the experimental value $M_{\rm top}^{exp}
= (170.9 \pm 1.8)$~GeV~\cite{mt1709}.  For the bottom quark mass we
used the running mass evaluated at $M_z$ $m_{\rm bot}(M_Z) = 2.82 \pm
0.07$~\cite{PDG} to avoid the uncertainties from the running of $M_Z$
to the $m_b$ pole mass, which are not related to the predictions of
the FUT models.  

\begin{figure}
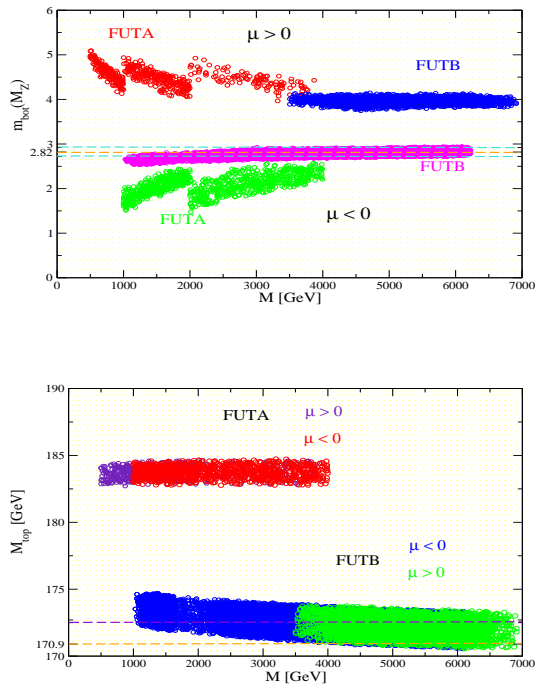

\vspace{0.5cm}
           \centerline{\includegraphics[width=7cm,height=4cm,angle=0]{MvsMBOT.eps}}
\vspace{1cm}
           \centerline{\includegraphics[width=7cm,height=4cm,angle=0]{MvsMTOP.eps}}
\vspace{0.5cm}
       \caption{The bottom quark mass at the $Z$~boson scale (upper) 
                and top quark pole mass (lower plot) are shown 
                as function of $M$ for both models.}
\label{fig:MtopbotvsM}
\vspace{-0.5em}
\end{figure}

In fig.\ref{fig:MtopbotvsM} we show the {\bf FUTA} and {\bf FUTB}
predictions for $M_{\rm top}$ and $m_{\rm bot}(M_Z)$ as a function of
the unified gaugino mass $M$, for the two cases $\mu <0$ and $\mu >0$.
In the value of the bottom mass $m_{\rm bot}$, we have included the
corrections coming from bottom squark-gluino loops and top
squark-chargino loops~\cite{deltab}, known usually as the $\Delta_b$
effects.  The bounds on the $m_{\rm bot}(M_Z)$ and the $M_{\rm top}$
mass clearly single out \FUTB\ with $\mu <0$, as the solution most
compatible with this experimental constraints.
 Although $\mu < 0$ is already challenged by present data of
the anomalous magnetic moment of the muon $a_{\mu}$, a heavy SUSY
spectrum as the one we have here gives results for $a_{\mu}$ very
close to the SM result, and thus cannot be excluded on this fact
alone.
 
In addition the value of $\tan \beta$ is found to be $\tan \beta \sim 54$ and
$\sim 48$ for models ${\bf A}$ and ${\bf B}$, respectively.
Thus the comparison of the model predictions
with the experimental data is survived only by {\bf FUTB} with $\mu < 0$.

We now analyze the impact of further low-energy observables on the model
{\bf FUTB} with $\mu < 0$.
As  additional constraints we consider the following observables: 
the rare $b$~decays $\br(b \to s \gamma)$ and $\br(B_s \to \mu^+ \mu^-)$, 
the lightest Higgs boson mass 
as well as a loose CDM constraint, assuming it
consists mainly of neutralinos. More details and a complete set of
references can be found in ref.~\cite{Heinemeyer:2007tz}.

For the branching ratio $\br(b \to s \gamma)$, we take
the present 
experimental value estimated by the Heavy Flavour Averaging
Group (HFAG) is~\cite{bsgexp} 
\beq 
\br(b \to s \gamma ) = (3.55 \pm 0.24 {}^{+0.09}_{-0.10} \pm 0.03) 
                       \times 10^{-4}.
\label{bsgaexp}
\eeq
For the branching ratio $\br(B_s \to \mu^+ \mu^-)$, the SM prediction is
at the level of $10^{-9}$, while the present
experimental upper limit from the Tevatron is 
$5.8 \times 10^{-8}$ at the $95\%$ C.L.~\cite{bsmmexp}, providing the
possibility for the MSSM to dominate the SM contribution.

Concerning the lightest Higgs boson mass, $M_h$, the SM bound of
$114.4$~GeV~\cite{LEPHiggsSM_MSSM} can be used. For the
prediction we use the code 
{\tt FeynHiggs}~\cite{feynhiggs}. 

The lightest supersymmetric particle (LSP) is an
excellent candidate for cold dark matter (CDM)~\cite{EHNOS}, with a density 
that falls naturally within the range 
\beq
0.094 < \Omega_{\rm CDM} h^2 < 0.129
\label{cdmexp}
\eeq favoured by a joint analysis of WMAP and other astrophysical and
cosmological data~\cite{WMAP}.  Assuming that the cold dark matter is
composed predominantly of LSPs, the determination of $\Omega_{\rm CDM}
h^2$ imposes very strong constraints on the MSSM parameter space, and
we find that no FUT model points fulfill the strict bound of
\ref{cdmexp}. On the other hand, many model parameters would yield a
very large value of $\Omega_{\rm CDM}$.  It should be kept in mind
that somewhat larger values might be allowed due to possible
uncertainties in the determination of the SUSY spectrum (as they might
arise at large $\tb$, see below).  Therefore, in order to get an
impression of the {\em possible} impact of the CDM abundance on the
collider phenomenology in our model, we will
analyze the case that the LSP does contribute to the CDM density, and
apply a more loose bound of \beq \Omega_{\rm CDM} h^2 < 0.3~.
\label{cdmloose}
\eeq
 Notice that lower values than the ones permitted by (\ref{cdmexp})
are naturally allowed if another particle than the lightest neutralino
constitutes CDM.  For our evaluation we have used the code
{MicroMegas}~\cite{micromegas}.

\begin{figure}
           \centerline{\includegraphics[width=7cm,angle=0]{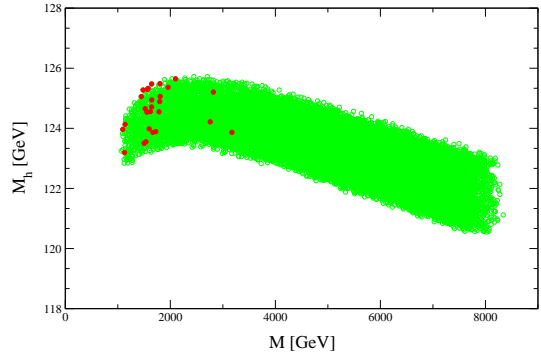}}
        \caption{The lightest Higgs mass, $M_h$,  as function of $M$ for
          the model {\bf FUTB} with $\mu < 0$, see text.}
\label{fig:Higgs}
\vspace{-0.5em}
\end{figure}

The prediction for $M_h$ of {\bf FUTB} with $\mu < 0$ is shown in
Fig.~\ref{fig:Higgs}. 
The constraints from the two $B$~physics observables are taken into
account. In addition the CDM constraint (evaluated with 
{\tt Micromegas}~\cite{micromegas}) is fulfilled for the darker
(red) points in the plot, see ref.~\cite{Heinemeyer:2007tz} for details. 
The lightest Higgs mass ranges in 
\beq
M_h \sim 121-126~{\rm GeV} , 
\label{eq:Mhpred}
\eeq 
where the uncertainty comes from
variations of the soft scalar masses, and
from finite (i.e.~not logarithmically divergent) corrections in
changing renormalization scheme.  To this value one has to add $\pm 3$
GeV coming from unkonwn higher order corrections~\cite{mhiggsAEC}. 
We have also included a small variation,
due to threshold corrections at the GUT scale, of up to $5 \%$ of the
FUT boundary conditions.  
Thus, taking into account the $B$~physics constraints (and possibly the
CDM constraints) results naturally in a light Higgs boson that fulfills
the LEP bounds~\cite{LEPHiggsSM_MSSM}. 

\begin{figure}
           \centerline{\includegraphics[width=10cm,angle=0]{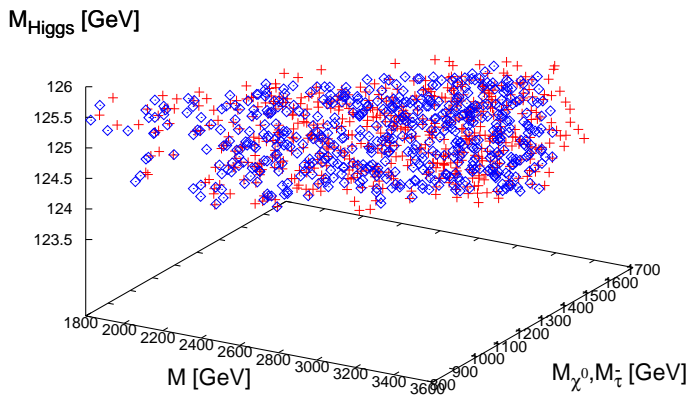}}
           \caption{The lightest Higgs mass, $M_h$, plotted against $M$
             and the LSP, which can be the neutralino $\chi^0$ (red
             crosses) or the stau $\tilde\tau$ (blue squares), for
             the model {\bf FUTB} with $\mu < 0$, see text.}
\label{fig:3d}
\vspace{-0.5em}
\end{figure}

In Fig.~\ref{fig:3d} we present the Higgs mass for \FUTB\ for the case
when the LSP is the neutralino $\chi^0$ (red crosses) and when it is
the stau $\tilde \tau$ (blue squares), for the range of values of the
gaugino mass $M$ where the loose CDM constraint is fulfilled (left
part of Fig.~\ref{fig:Higgs}).  From Fig.~\ref{fig:3d} it is clear
that the prediction for the Higgs mass lies in the same range for both
cases.  Notice that in case the LSP is the s-tau it can decay by
introducing bilinear R-parity violating terms, which respect the
finiteness conditions.  R-parity violation would have a small impact
on the collider phenomenology presented here, but would remove the CDM
bound (\ref {cdmexp}) completely and the LSP would not be the CDM
candidate.

In the same way the whole SUSY particle spectrum can be derived. 
The resulting SUSY masses for {\bf FUTB} with $\mu < 0$ are rather large.
The lightest SUSY particle starts around 500 GeV, with the rest
of the spectrum being very heavy. The observation of SUSY particles at
the LHC or the ILC will only be possible in very favorable parts of the
parameter space. For most parameter combinations only a SM-like light
Higgs boson in the range of eq.~(\ref{eq:Mhpred}) can be observed.

\bigskip
 Partially supported by the NTUA programme for basic research
``K. Karatheodoris''.  Work supported in part by EU's Marie-Curie
Research Training Network under contract MRTN-CT-2006-035505 `Tools
and Precision Calculations for Physics Discoveries at Colliders'.
Also supported by the mexican grants PAPIIT-UNAM IN115207 and Conacyt
51554-F.

\end{document}